\documentclass[preprint,10pt]{elsarticle}

\usepackage{amssymb}

 \usepackage{lineno}

\biboptions{square,comma,sort&compress, numbers}

\begin{document}

\begin{frontmatter}



\title{Comments on: {\it Measurement of Parton Distributions of Strange Quarks in the Nucleon
from Charged-Kaon Production in Deep-Inelastic Scattering on the Deuteron} by the HERMES 
Collaboration}


\author{M. Stolarski}

\address{LIP, Av. Elias Garcia 14 - 1º 1000-149 Lisboa, Portugal}

\begin{abstract}

In this paper a discussion is made of the article by the HERMES Collaboration, 
Phys.~Lett.~B666,~446 \cite{her1}, where
several important results concerning strange quark properties in the nucleon were presented. 
By analysing the sum of $K^\pm$ multiplicities it was found that 
the strange sea distribution is very different from the non-strange one
as a function of the Bjorken $x$ scaling variable. 
In addition, the magnitude of these two distributions at low $x$ is similar, 
contrary to the behaviour in most of the available parton distribution function sets.

It is shown that the obtained distribution of the unpolarised strange quark
influences our understanding of the ``strange quark polarisation puzzle''.
It is argued that the sole analysis of the sum of $K^{\pm}$ multiplicities,
as done in the HERMES paper, might not be sufficient to draw 
solid conclusions about the strange sector.
A simultaneous analysis of the difference of $K^{\pm}$ multiplicities 
should be done in general. 
To show that this is not simply an academic case, the author presents an analysis based on the HERMES
preliminary data, where the sum and the difference of the $K^{\pm}$ multiplicities
are considered.

\end{abstract}


\end{frontmatter}



\section{Introduction} \label{sec:int}

The information on the strange quark properties 
in the nucleon is rather scarce. 
The strange quark density is poorly
known; the main sources of information are the neutrino
experiments NuTeV \cite{nutev} and CCFR \cite{ccfr}. The analyses 
\cite{mstw, abkm, ball1, ball2}
suggest that the strange sea is suppressed with
respect to the non-strange sea by a factor of about 2. 
This fact is not observed in the analysis of \cite{lai}.

The HERMES results \cite{her1} brought a very important contribution 
to the understanding of the strange quark properties in the nucleon. 
First of all it was shown that the strange
quark parton distribution function (PDF) has very different shape 
in the Bjorken $x$ scaling variable as compared to the corresponding
distributions of $\bar{u}$ and $\bar{d}$.
It was also verified that for $x \approx 0.04$ 
the densities of the strange and non-strange sea are comparable. 
This result was recently confirmed by the ATLAS analysis \cite{atlas1}, 
but the final uncertainty of the ATLAS result is still large. 

Significant results were also obtained by HERMES 
in the polarised strange sector. 
From deep inelastic scattering measurements of the spin dependent
structure function $g_1$ it is known that the first moment
of the strange quark distribution is negative $\Delta S= \int_0^1 \Delta s(x) + \Delta \bar{s}(x) dx= 
-0.09 \pm 0.01 \pm 0.01$,
(see $e.g.$ \cite{h2}, \cite{qcd_compass}).
However, HERMES analysis based on the combination of the inclusive asymmetry $A_{||,d}$
and the semi-inclusive kaon asymmetry $A_{||,d}^{K}$ concluded that, in the measured
range of $x$, the corresponding $\Delta S$ is consistent with zero and negative values
are not preferred: $\int_{0.02}^{0.6}\Delta s(x)+ \Delta \bar{s}(x) dx= 0.037 \pm 0.019 \pm 
0.027$. 

The HERMES observation lead to a sign changing solution of the polarised 
strange quarks distribution function in a PDF set by
D. de Florian, R. Sassot, M. Stratmann and W. Vogelsang \cite{dssv}. 
In the high $x$ region the strange quark polarisation is positive, 
changing sign for  low values of $x$.
As a result the semi-inclusive and the inclusive analyses of $\Delta S$ 
no longer contradict each other.

On the other hand 
E. Leader, A. V. Sidorov and D. B. Stamenov in Ref. \cite{lss}
argue that the NLO fits of inclusive (and even semi-inclusive $\pi^{\pm}$)
data lead to a negative solution for $\Delta S$ in the {\bf whole} $x$ range.
Only when the kaon asymmetries are included in the fit, the sign 
changing solution appears. They call this the ``strange polarisation puzzle''.
They further argue that the puzzle only appears if one uses the DSS set of fragmentation
functions (FFs)  \cite{dss}, as done in \cite{her1}. 
Using another set, namely the HKNS \cite{hkns} one, also kaon semi-inclusive data 
prefer negative strange polarisation in the whole $x$ range. 

Based on published materials in \cite{her1} 
it can be shown that the proposition of \cite{lss}
has a problem.
The strange quark fragmentation function into kaons is more than 
a factor two lower in HKNS than in DSS. 
HERMES used in \cite{her1} the DSS FF set in their extraction of $S(x)=s(x)+\bar{s}(x)$,
obtaining $S(x)$ as large as $\bar u+ \bar d$ in the low $x$ region.
This large $S(x)$ came rather as a surprise. 
But if HERMES had used the HKNS FF set for the extraction of $S(x)$
the resulting distribution would have been more than a factor two higher 
than  $\bar u+ \bar d$ in the low $x$ region, and in disagreement with all 
available experimental data. 
Formally the solution proposed in \cite{lss} solves the puzzle in the polarised
strange sector, but creates another one for the unpolarised case.

However, in this paper it is argued that the conclusions 
of the HERMES paper concerning $S(x)$ might be premature. 
The reason is that in the HERMES analysis only the sum of $K^\pm$ multiplicities
was included in the analysis, whereas  
the simultaneous analysis
of the difference of $K^{\pm}$ multiplicities should be also studied.
This is especially important when some unexpected behaviour is found in the analysis
of the sum of $K^\pm$ multiplicities.
To support further these arguments, the results of the analysis
of the HERMES preliminary  data \cite{h_dis11} will be shown.

\section{HERMES analysis of unpolarised strange sea}

The HERMES results \cite{her1} on the strange quark PDF $S(x)$ are based on the
analysis of the sum of $K^\pm$ multiplicities $dN^K(x)/dN^{DIS}(x)$ in semi-inclusive deep 
inelastic 
scattering (SIDIS) of electrons impinging on a deuteron target.  
As stated there,

\begin{equation}
S(x) \int D_S^K(z) dz \simeq  Q(x) \left [5 \frac{d^2N^K(x)}{d^2N^{DIS}(x)} - \int D_Q^K(z)dz \right ],
\label{eq:1}
\end{equation}
\noindent
where $z$ is the ratio of energies of the hadron and the virtual
photon in the target rest frame,
$D_Q^K(z) \equiv 4D_{u+\bar{u}}^K(z)+ D_{d+\bar{d}}^K(z)$ and
$D_S^K(z) \equiv 2D_{s+\bar{s}}^K(z)$.

As presented in Fig.~1 of \cite{her1}, the kaon multiplicity is flat for high $x$,
$i.e.$ in the region where there is no strange quarks and it rises by about
50\% for lower values of $x$.
Without strange quarks the distribution should be almost flat; 
therefore the large excess of kaon multiplicity in the low $x$ region
is interpreted as a strong signature of the strange quarks presence. 
Their contribution was quantified using Eq.~(\ref{eq:1}). 
The value of $\int D_S^K(z) dz$ was taken from DSS
and $\int D_Q^K(z)dz$ was extracted directly from the HERMES data at high $x$. 
At first order, 
the HERMES analysis neglected the possible negative four momentum transfer $Q^2$ dependence of the FF. 
However, in the later stage of the analysis the dependence as in the DSS 
parametrisation of the fragmentation 
functions was taken
into account. The expected $Q^2$ dependence of $D_Q^K$ is weak, of the order of 5\%, in the measured 
range of $Q^2\in (1-10)$ (GeV/c)$^2$.

\section{ The $K^+ - K^-$ multiplicity difference and $D_Q^K$}

To be able to perform a quantitative analysis,
let us assume that
in the strange sector there are three fragmentation functions:
$D_{str}= D_s^{K^{-}} = D_{\bar{s}}^{K^{+}}$;
$D_{fav}= D_{\bar{u}}^{K^{-}}= D_{u}^{K^{+}}$; and $D_{unf}$ for the remaining combinations.
Thus: $D_Q^K \equiv 4D_{u+\bar{u}}^K+ D_{d+\bar{d}}^K= 4 D_{fav}+ 6 D_{unf}$ and
$D_S^K \equiv 2D_{s+\bar{s}}^K = 2 D_{str}+ 2 D_{unf}$.
Here for simplicity it is assumed that $D_i \equiv \int_{0.2}^{0.8}D_i(z)dz$.

An important distribution to understand better the strange sector
is the difference of $K^+$ and $K^-$ multiplicities, $dN_{diff}^K(x)/dN^{DIS}(x)$.
A lot of systematic uncertainties cancel in the extraction of the multiplicity difference.
Also  the gluon contribution
cancels, which leads to a simpler evolution in NLO.
However, most importantly the strange quarks contribution
{\bf cancels} in the kaon multiplicity difference. 
Thus an analysis of the multiplicity difference is simpler, both experimentally and theoretically,
than a separate analysis of $K^+$ and $K^-$ multiplicities or their sum.
In LO using Eq.~(\ref{eq:my1}) one has a direct link to a certain combination of non-strange FFs, namely
$D_{fav} - D_{unf}$ :
\begin{equation}
\frac{dN_{diff}^{K}}{dN^{DIS}}=  \frac{4(u_v+ d_v)}{ 5( u+\bar{u}+d+\bar{d})+4s} (D_{fav} - D_{unf})
\label{eq:my1}
\end{equation}
Equation (\ref{eq:my1}) applies for the deuteron target, where for simplicity the $x$ and $Q^2$ 
dependencies are omitted.
The unpolarised quark distributions of various flavors
are denoted by $u$, $\bar{u}$, $d$, $\bar{d}$, $s$.
The denominator of the equation is closely related to the measured single photon exchange cross-section,
while $u_v$ and $d_v$ are the valence distributions.
Both factors are known experimentally with a good precision.
Hereafter the $\int_{0.2}^{0.8} D_{fav} - D_{unf}(z)dz$ will be noted as $D_{F-U}$.

The  study of the kaon production irrespective of charge using an isoscalar target
indeed simplifies the analysis of \cite{her1}. It reduces uncertainties related 
to the detailed description of  the different fragmentation functions,
$e.g.$ a separate knowledge of $D_{fav}$ and $D_{unf}$ is not necessary.
Therefore it may seem that 
the $dN_{diff}^K(x)/dN^{DIS}(x)$  is not related to the HERMES analysis of $dN^K(x)/dN^{DIS}(x)$. 
However, according to the author this is only true when 
$dN_{diff}^K(x)/dN^{DIS}(x)$ is well under control. 
In other words if some peculiarities are observed in $dN_{diff}^K(x)/dN^{DIS}(x)$
they can influence the observed $dN^K(x)/dN^{DIS}(x)$. At the same time
it is known that they cannot be related to the strange quarks, since 
their contribution cancels in $dN_{diff}^K(x)/dN^{DIS}(x)$.

For simplicity, let us consider the region of high $z$, where one
can safely assume that only $\bar{s}$ and $u$ contribute
to the $K^+$ production and their charge conjugate
for the $K^-$.
Furthermore let us assume that the measurement of $K^\pm$ 
was performed in two regions of $x$;
$x_{high}$,
where there is no $s, \bar{s}$ contribution and $x_{low}$ where 
all four considered flavours contribute. 
Finally let us assume that there is a difference observed between $D_{F-U}$ in these two $x$ regions.

In such a simplified case $D_Q^K$ 
is directly proportional to $D_{F-U}$;  $D_Q^K = 4D_{F-U}$.
Thus it is clear that the observed dependence 
$D_{F-U}$ between $x_{low}$ and $x_{high}$ will have a direct influence 
on the obtained $D_Q^K$ in these two regions, and so
extracted $S(x) \int D_{str}(z)  dz$,  using Eq.~(\ref{eq:1}), is also affected. 
Moreover in the described situation one can obtain
correct results of $S(x) \int D_{str}(z)  dz$ using information only from the low $x$ region.
On the other hand  neglecting the information from $dN_{diff}^K/dN^{DIS}(x_{low})$, 
and using $D_Q^K(x_{high})$ instead, leads to a bias of
the extracted $S(x) \int D_{str}(z) dz$. 

In the more general case $D_Q^K = 4 D_{F-U} + 10 D_{unf}$, 
a strong $x$ or $Q^2$ dependence of $D_{F-U}$ can influence
$D_Q^K$~\footnote{Observe that this conclusion
is valid even in the  case when one assumes only the charge conjunction asymmetry of the FFs, $i.e.$
6 independent FFs in the strange quark sector.}~.
However,  one has to make additional assumptions
on which FF(s) contribute to the observed $x$ dependence of $D_{F-U}$.
Let us assume that we observe $dN^K(x)/dN^{DIS}(x)$ as in \cite{her1},
but also that $D_{F-U}(x_{low})$ is somewhat larger than
$D_{F-U}(x_{high})$.
The impact of the $x$ dependence of $D_{F-U}$ on the extracted $S(x) \int D_{str}(z)dz$
is studied in two extreme scenarios where only either $D_{unf}$ or $D_{fav}$ is responsible
for the observed $x$ dependence of $D_{F-U}$.

In the first scenario only $D_{unf}$ is responsible
for the $x$ dependence  of $D_{F-U}$.
Effectively, as $D_Q^K(x_{low})$ is now lower 
than $D_Q^K(x_{high})$, the value of 
$S(x) \int D_{str}(z)dz$ is  increased. 
However, $D_{unf}=0.008$ in DSS for $Q^2=1$ (GeV/c)$^2$; thus there
is not much room left to reduce its value further. 

In the second scenario $D_{fav}$ increases for low $x$. 
This would generate an  increase of the sum of $K^{\pm}$ multiplicities
for low $x$ but originating from $u$, $\bar{u}$ instead of $s$, $\bar{s}$.
As a result the extracted $S(x) \int D_{str}(z) dz$ would be lower. 
Moreover assuming that $S(x)$ is known, as done at the beginning of the HERMES analysis,
the second scenario could lead to lower values of the extracted $D_{str}$.
Observe that as $D_{str}$ is lower, the errors of the extracted $\Delta S(x)$
increase (see $e.g.$ \cite{compass_fl}).
On top of that, more positively polarised $u$ quarks
contribute to the kaon sample from which the spin dependent asymmetry $A_{||,d}^K$ is extracted.
Failing to take this into account would lead to a bias of  $\Delta S$ towards positive values.

Unfortunately in the general case the observed difference of $D_{F-U}$
cannot be attributed easily to one of the two functions. In such a case the uncertainty
of the extracted $S(x)\int D_{str}(z) dz$ is largerly increased.
In the author's opinion this means that without a proper understanding of the behaviour 
of the difference of $K^{\pm}$ multiplicities, one cannot really reach a solid conclusion
concerning  $S(x)\int D_{str}(z) dz$.
If some peculiarities are observed in the sum of $K^{\pm}$ multiplicities
one cannot analyse just these, as done in \cite{her1}, but the simultaneous
analysis of the multiplicities difference in such case is mandatory. 
The analysis as in \cite{her1} is fully justified only when 
the kaon multiplicities difference is well understood.

\section{Tentative analysis of the  HERMES publicly available data}

The previously stated claims cannot be confronted with published data,
since the Collaboration published only the sum of the $K^{\pm}$ multiplicities.
Instead preliminary HERMES data are used here.
Since the data are only preliminary, the results
and conclusions should be considered only as a {\em proof of principle}
of what was said in the previous section.

The $K^+$ and $K^-$ multiplicities from deuteron target were extracted
using the data available in \cite{h_dis11}, slide 15. 
These multiplicities cover the following kinematic ranges: $x \in (0.03-0.5)$ and $Q^2 \in (1-10)$ (GeV/c)$^2$.
They were integrated in the $z$ range from 0.2 to 0.8.
The extracted values are summarised in Tables~\ref{tab:1} and \ref{tab:2} of the Appendix. 
In all but the last $x$ interval only the systematic uncertainty could be extracted, as 
the statistical error is smaller than the size of the point on the figures.  
Although the systematic errors of $K^+$ and $K^-$ multiplicities in a given $x$ interval
are known to be strongly correlated, 
the covariance matrix of the systematic errors is not available to the author
at this stage of the analysis.
Based on the behaviour of the systematic errors for $K^+$ and $K^-$  multiplicities
the correlation factor was estimated to be about 0.8 .

The sum of the extracted $K^{\pm}$ multiplicities at $x>0.1$ is about 0.10,
which is 25\% higher than 0.08 in \cite{her1}. The behaviour at lower
$x$ is very similar to the one in the discussed paper, just the increase of the multiplicities
between the low and high $x$ ranges is 30\% instead of 50\%.  
This observation could not be verified
because the author has not found any comparison, performed by the HERMES Collaboration,
between the new preliminary data and the published in \cite{her1}. 
The above difference between the preliminary
and the published results does not have an impact on the manuscript main conclusions.

\begin{figure}[!ht]
\begin{center}
\includegraphics[width=0.60\textwidth]{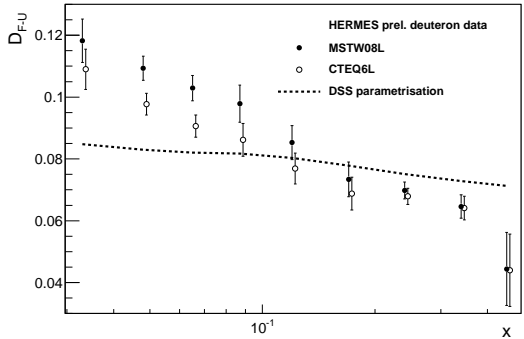}
\caption{The distribution of $D_{F-U}$ as a function of
$x$, obtained from HERMES preliminary data \protect{\cite{h_dis11}}.
The closed (open) points correspond to an analysis where the MSTW08L (CTEQ6L) PDF set was 
used.
The dashed line corresponds to the DSS predictions
related with the $Q^2$ dependence of the fragmentation functions.} 
\label{fig:1}
\end{center}
\end{figure}

An evaluation of $D_{F-U}(x)$ was made, 
using Eq.~(\ref{eq:my1}) and the CTEQ6L \cite{cteq} and MSTW08L \cite{mstw} PDF sets.
The MSTW08L set is more recent, and in addition in CTEQ6L the
$Q^2$ evolution is frozen below $Q^2=1.69$ (GeV/c)$^2$.
However, since the CTEQ6L was used in the original HERMES work the analysis
was performed using both PDF sets. The quark densities used in the analysis are
summarised in Tables~\ref{tab:3} and \ref{tab:4} of the Appendix. 
The resulting  $D_{F-U}(x)$  are presented in Fig.~\ref{fig:1}.
The points are the $D_{F-U}(x)$ values, extracted using MSTW08L (closed) and CTEQ6L (open points).
The dashed-line is the DSS parametrisation, taking into account the average
$Q^2$ values in the different $x$ intervals.
The observed $x$ or $Q^2$ dependence of the evaluated  $D_{F-U}$ is 
stronger than expected from DSS, especially when using the MSTW08L PDF set.

\begin{figure}[!ht]
\begin{center}
\includegraphics[width=1.0\textwidth]{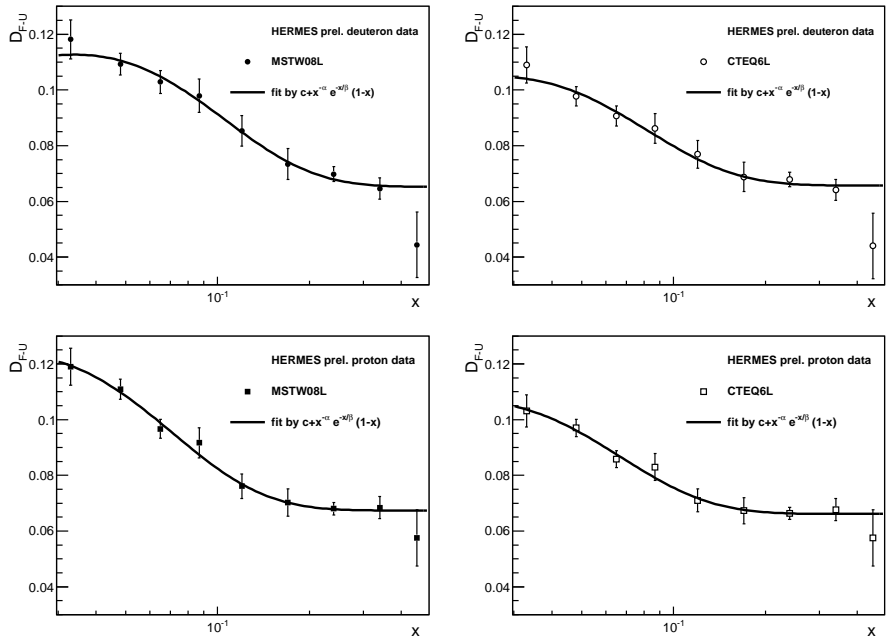}      
\caption{The distribution of $D_{F-U}$ as a function of $x$,
obtained in the present analysis of the HERMES preliminary data \protect{\cite{h_dis11}}.
In the left (right) panels the extraction 
for MSTW08L (CTEQ6L) PDF sets are shown.
In the upper (lower) panels data obtained with a deuteron (proton) targets
are shown. 
The continuous line is a fit using the functional form $c+x^{-\alpha} e^{-x/\beta}(1-x)$.}
\label{fig:2}
\end{center}
\end{figure}

The natural question is to what extent $x$ or $Q^2$ dependence of $D_{fav}$
could explain simultaneously the observed features of the sum and the difference of $K^{\pm}$ 
multiplicities~\footnote{Here it is 
assumed that the observed dependence has a physical origin, while in principle the results
could suggest also problems in the multiplicity extraction method for the preliminary data.}~.  
The HERMES paper concluded that their assumed $x S(x)$ dependence could be fitted by the functional form
$x^{-\alpha} e^{-x/\beta}(1-x)$. In the considered $Q^2$ range and for  $x\in (0.03,0.17)$ 
the $x Q(x)$ changes only by 10\%.
In such a case the possible $x$ dependence of $D_{fav} \rightarrow D_Q^K$ 
is directly proportional
to the bias of the extracted $x S(x)$, see Eq.~(\ref{eq:1}).
Therefore it is natural to use the same functional form to describe 
the $x$ dependence of 
$D_{F-U}$ as was used by HERMES to describe $xS(x)$~\footnote{This should be considered as an effective 
description, without a physical sense for $D_{F-U}$ outside of the fitted range.}~.
The results of the fit for both PDF sets are presented in the upper panels of Fig.~\ref{fig:2};
MSTW08L and CTEQ6L on the left and right panels respectively. 
In addition also the 
$D_{F-U}(x)$ were extracted using the proton data shown in \cite{h_dis11}. The results and
the fits are presented in the lower panels of Fig.~\ref{fig:2}.  
There is a good agreement between the results extracted for
the proton and deuteron cases. 
However, since the present analysis is mostly concentrated on the deuteron data,
the fit parameters for proton and deuteron were kept independent.

The difference between parametrised $D_{F-U}(0.033)-D_{F-U}(0.45)$,
called $D_{F-U}^{low-high}$ is equal to $0.047\pm 0.004$ and $0.038 \pm 0.004$
using the MSTW08L and CTEQ6L PDFs respectively. 
It was also verified that a similar, but weaker, dependence is also present
in the results from the PhD thesis of \cite{h_phd} \footnote{The results of that thesis were used
in the DSS parametrisation of FFs.}, (Tables B.7 and B.9).
In this case $D_{F-U}^{low-high}$= 0.030 and 0.024 for MSTW08L and
CTEQ6L respectively. 
However, for those results  the multiplicities were evolved to the same $Q^2$ value,
thus the distribution should be flat. 
In addition a region in $z$ smaller than in \cite{her1} was available.
In \cite{h_phd} it is mentioned that there is possibly a problem in the procedure
of the multiplicity extraction, since the extracted $K^-$ multiplicities were sometimes negative.

The verification of a possible impact of the observed
$D_{F-U}(x)$ dependence on the $D_{str}$ (one could also study the 
impact on $S(x)\int D_{str}(z)dz$ instead) follows. 
A simple LO extraction of FFs 
from HERMES data was performed using as input the extracted integrated 
$K^+$ and $K^-$ multiplicities from \cite{h_dis11} and two PDF sets. 
The correlation factor between the systematic errors of $K^+$ and $K^-$ 
was not taken into account. The main reason is that one should know
the full covariance matrix while performing the fit.

Without any $Q^2$ dependence of FFs the fit results are summarised in Table~\ref{tab:res:1}. 
The $\chi^2/ndf$ are very high, thus the data cannot be describe in such a fit.
The conclusion is the same as in the discussed paper.
For comparison, the results of the LO DSS analysis of the world data at $Q^2$= 2.5 (GeV/c)$^2$ are
also shown in the table. The obtained results are not so different from those.

\begin{table}[h]
\begin{center}
\caption{The results of the fit to the HERMES preliminary data using constant FFs. For comparison,
the results of the world data fit of DSS are also given.  }
\label{tab:res:1}
\begin{tabular}{c|c|c||c}
  &using MSTW08L & using CTEQ6L & DSS \\ \hline
$D_{fav}$    & $0.100 \pm 0.003$ & $0.096 \pm 0.003$ &$ 0.091$ \\ 
$D_{unf}$    & $0.017 \pm 0.002$ & $0.018 \pm 0.002$ &$ 0.012$ \\ 
$D_{str}$    & $0.45 \pm 0.09  $ & $0.50 \pm 0.09  $ &$ 0.62 $ \\ 
$\chi^2/ndf$ & $75.4/15        $ & $57.1 /15       $ &$  -    $ \\ 
\end{tabular}
\end{center}
\end{table}

At this stage of the analysis in the HERMES paper it was concluded 
that the bad description of the data by the fit 
is related to the usage of an incorrect $S(x)$ distribution. 
In the following steps of their analysis 
the $S(x)$ was extracted from the data, while the $D_{str}$ from DSS was used.

Formally one could describe the preliminary HERMES in such a way. 
Indeed the sum of kaon multiplicities is then well described.
However, the $D_{F-U}$ is not affected by the $S(x)$ change, since the
contribution of strange quarks cancels in the multiplicity difference. Thus the
unexpected behaviour of $D_{F-U}$ observed in Fig.~\ref{fig:1} anyhow has to be
explained by a $x$ or $Q^2$  dependencies of $D_{fav}$ and $D_{unf}$.
Moreover, it cannot be any dependence but a fine tuning of the parameters is needed.
The unexpected dependencies of $D_{fav}$ and $D_{unf}$ have to cancel in  $D_Q^K(x)$,
as otherwise the $S(x)$ extracted from the multiplicity sum will be biased.

The author does not follow the above solution, but assumes that the $D_{fav}$ changes 
as: $D_{fav}(x)=D_{fav}(1)+ x^{-\alpha} exp(-x/\beta)(1-x)$, as motivated earlier.    
At the same time $S(x)$ is still taken from the corresponding PDF set. 
The results of such a fit are summarised in Table~\ref{tab:res:2}.
Comparing to the previous results one sees that the $\chi^2/ndf$ of the fit improved by a factor 7-8,
and the data are well described by the assumed functional form. 
However, as expected, there are large changes observed for $D_{str}$.
In fact for MSTW08L the extracted value of $D_{str}=-0.48 \pm 0.15$
is unphysical. The reason is that the $D_{F-U}^{low-high}$ is so large
that it overshoots the multiplicity sum in the simplest scenario considered.
The situation is better for CTEQ6L, where the extracted value 
of  $D_{str}=-0.25 \pm 0.15$ has an overlap with the physically allowed region
within $3\sigma$.
It was also verified that including a $Q^2$ dependence of $D_{unf}$ proportional to 
the DSS results, the values of $D_{str}$ are increased by about 0.2-0.25. In such case the
overlap with a physically allowed region is largely increased.

However, the main point of this paper is to show that in case we do not understand
the $D_{F-U}(x)$ dependence it is very hard to extract reliable information about 
the $D_{str}$ or more generally about $S(x) \int D_{str}(z) dz$. 
In the exercise above it was assumed that the $x$ dependence of $D_{F-U}$
was related to $D_{fav}$. 
However, this is only an assumption. 
One should consider also the possibility of a $D_{unf}$ change.
Again, for simplicity, the author assumed that:
$\begin{array}{lcl}
D_{fav} &=& D_{fav}(1)+ P_{fav} \; x^{-\alpha} exp(-x/\beta)(1-x) \\ 
D_{unf} &=& D_{unf}(1)+ P_{unf} \; x^{-\alpha} exp(-x/\beta)(1-x) \\
\end{array}$ \\
So, $\alpha$ and $\beta$ parameters are the same in both cases. 
The $\chi^2/ndf$ of the fits using different PDF sets improved to about 5.6/11.
However, even in this rather simple case, the parameters have such a large correlation that 
the uncertainty of $D_{str}$ is in the range 0.7-0.8. 
This confirms that data containing unforeseen  $D_{F-U}(x)$ dependence, 
can hardly be a reliable source of information about $D_{str}$ or $S(x) \int D_{str}(z) dz$.

\begin{table}
\begin{center}
\caption{The results of the fit to the HERMES prel. data using $D_{fav}(x)= D_{fav}(1)+x^{-\alpha} exp(-x/\beta)(1-x)$ and
constant $D_{unf}$ and $D_{str}$}. 
\label{tab:res:2} 
\begin{tabular}{c|c|c}
  &using MSTW08L & using CTEQ6L  \\ \hline
$D_{fav}$    & $0.093 \pm 0.003$ & $0.092 \pm 0.003$   \\ 
$D_{unf}$    & $0.027 \pm 0.002$ & $0.027 \pm 0.002$   \\ 
$D_{str}$    & $-0.48 \pm 0.15  $ & $-0.25 \pm 0.15  $   \\ 
$\alpha$     & $ -0.57\pm 0.04  $ & $-0.59 \pm 0.06  $  \\ 
$\beta $     & $ 0.039 \pm 0.004$ & $0.033 \pm 0.005 $  \\ 
$\chi^2/ndf$ & $9.7/13        $ & $8.7 /13$        \\ 
\end{tabular}
\end{center}
\end{table}

\section{Miscellaneous}

As stated in the HERMES paper 
the value of $\Delta S$ extracted from the analysis of kaon asymmetries
is not necessarily in serious disagreement with the value obtained in the inclusive
world data fit, since the two estimates cover different $x$ ranges.
The two values are $\int_{0.02}^{0.6} \Delta S(x) dx = 0.037 \pm 0.019 \pm 0.027$ extracted from
the kaon asymmetries and $\Delta S=-0.103 \pm 0.007 \pm 0.013 \pm 0.008$ for the inclusive case. 

However, take note that the assumed $S(x)$ distribution has an 
indirect impact on the analysis of $\Delta S$ performed in HERMES.
The results of the HERMES analysis of $x \Delta S(x)$ from \cite{her1} 
are summarised in the two first columns of Table \ref{tab:res:3}, where the values of $\langle x \rangle$ and
$x \Delta S(x)$ with statistical and systematic uncertainties are presented.
The following columns are the estimates done in the present analysis, for the bin width
and the contribution of the given bin to the integral of $ \int \Delta S(x) dx$.
It can be noticed that, within rounding errors, the value $0.037 \pm 0.019$ in
reality corresponds to the integral $\int_{0.02}^{0.14} \Delta S(x) dx$.
Thus most probably, in the HERMES analysis it was assumed that for $x>0.14$   the $S(x)$ distribution
vanishes, as obtained in their unpolarised analysis. The systematic error was instead increased
to take into account the extrapolation to the region $x \in (0.14-0.6)$. 
Now if the distribution of $S(x)$ assumed by HERMES
does not vanish so fast at high $x$,
two other $x$ points should be considered in the analysis of the $\Delta S$ integral.
This would lead to $\int_{0.02}^{0.30} \Delta S(x)=0.106 \pm 0.027 \pm 0.016$ or to
about $0.069 \pm 0.027 \pm 0.016$ if the positivity limit from MSTW08 is applied
on $\Delta S$. In any case the tension between $\Delta S$ obtained from kaon asymmetries and
from the inclusive analysis is further increased. It can be eliminated by a lower
$D_{str}$ as proposed  by the  LSS group \cite{lss}.

\begin{table}
\begin{center}
\caption{Results of the HERMES analysis of $\Delta S$.}
\label{tab:res:3}
\begin{tabular}{c|c|c|c}
$\langle x \rangle$ &  $x \Delta S(x)$ & bin width & $\int_{bin}  \Delta S(x) dx$ \\ \hline
0.033&$ 0.002 \pm  0.020 \pm  0.002$ & 0.020 & $0.0009\pm 0.0123\pm 0.0012$ \\
0.047&$ 0.038 \pm  0.019 \pm  0.007$ & 0.015 & $0.0121\pm 0.0059\pm 0.0022$ \\
0.065&$ 0.015 \pm  0.019 \pm  0.015$ & 0.020 & $0.0045\pm 0.0058\pm 0.0046$ \\
0.087&$ 0.037 \pm  0.024 \pm  0.009$ & 0.025 & $0.0105\pm 0.0068\pm 0.0026$ \\
0.118&$ 0.028 \pm  0.026 \pm  0.010$ & 0.040 & $0.0093\pm 0.0088\pm 0.0034$ \\
0.166&$ 0.123 \pm  0.032 \pm  0.029$ & 0.060 & $0.0444\pm 0.0116\pm 0.0105$ \\
0.239&$ 0.058 \pm  0.038 \pm  0.022$ & 0.100 & $0.0242\pm 0.0161\pm 0.0092$ \\
\end{tabular}
\end{center}
\end{table}

\section{Summary}

It was shown that an analysis based on the kaon multiplicities, 
claiming an unexpected feature observed
in the strange quark sector cannot only concentrate on the sum of the kaon multiplicities.
Instead a parallel analysis of the multiplicities difference is necessary,
even if an isoscalar target is being analysed. 
Thus the claims in Phys. Lett. B 666 (2008) 446
might be premature. 
An addendum to that paper should be considered, to include
the kaon multiplicities difference.
This would allow to verify that this variable 
does not show any unforeseen $x$ dependence.
Such verification is very important in order to better understand
the polarised and unpolarised strange quark sector 
in the fixed target experiments domain.

\section*{Acknowledgements}

The author would like to thank B. Stamenov for useful comments and discussion. 
This research was supported by the Portuguese Funda\c{c}\~ao para a Ci\^encia e 
Tecnologia, grant SFRH/BPD/64853/2009.

\section*{Appendix}

In order to make the verification of the presented results possible,
in Tables \ref{tab:1}-\ref{tab:4} the raw information
used in this analysis is given.

\begin{table}[h!]
\begin{center}
\caption{$K^+$ multiplicities extracted from \protect{\cite{h_dis11}}
and their errors. Different intervals of $x$ and $z$ are shown.
The final $z$-integrated multiplicity is presented in the last column.}
\label{tab:1}
\begin{tabular}{c|c|c|c|c|c|c}
 $\langle x \rangle $ &  $\langle Q^2 \rangle$  & $z \in (0.2-0.3)$&      $z\in(0.3-0.4)$&        $z \in (0.4-0.6)$&      
$z\in(0.6-0.8)$&  $\frac{dN^{K^{+}}(x)}{dN^{DIS}(x)}$ \\   \hline
0.033&  1.1& $0.279 \pm 0.021$ & $0.197 \pm 0.016$ & $ 0.116 \pm 0.007$ & $ 0.051 \pm 0.007$ & $  0.0810 \pm 0.0033$	  \\
0.048&  1.4& $0.265 \pm 0.010$ & $0.199 \pm 0.011$ & $ 0.112 \pm 0.005$ & $ 0.046 \pm 0.005$ & $  0.0780 \pm 0.0021$	  \\
0.065&  1.6& $0.260 \pm 0.019$ & $0.185 \pm 0.005$ & $ 0.112 \pm 0.005$ & $ 0.049 \pm 0.004$ & $  0.0765 \pm 0.0024$	  \\
0.087&  1.7& $0.222 \pm 0.029$ & $0.166 \pm 0.015$ & $ 0.112 \pm 0.003$ & $ 0.053 \pm 0.009$ & $  0.0718 \pm 0.0038$	  \\
0.120&  2.1& $0.231 \pm 0.036$ & $0.162 \pm 0.013$ & $ 0.104 \pm 0.005$ & $ 0.051 \pm 0.003$ & $  0.0704 \pm 0.0040$	  \\
0.170&  3.1& $0.226 \pm 0.044$ & $0.168 \pm 0.015$ & $ 0.107 \pm 0.005$ & $ 0.046 \pm 0.006$ & $  0.0700 \pm 0.0049$	  \\
0.240&  4.9& $0.250 \pm 0.023$ & $0.170 \pm 0.009$ & $ 0.107 \pm 0.005$ & $ 0.048 \pm 0.004$ & $  0.0731 \pm 0.0028$	  \\
0.340&  7.4& $0.249 \pm 0.040$ & $0.188 \pm 0.011$ & $ 0.118 \pm 0.007$ & $ 0.048 \pm 0.003$ & $  0.0769 \pm 0.0044$	  \\
0.450&  10.1&$0.293 \pm 0.069$ & $0.144 \pm 0.029$ & $ 0.101 \pm 0.004$ & $ 0.036 \pm 0.002$ & $  0.0712 \pm 0.0076$	  \\
\end{tabular}
\end{center}
\end{table}

\begin{table}[h!]
\begin{center}
\caption{$K^-$ multiplicities as in  Table \ref{tab:1}.} 
\label{tab:2}
\begin{tabular}{c|c|c|c|c|c|c}
 $\langle x \rangle$ &  $\langle Q^2 \rangle$  & $z \in (0.2-0.3)$&      $z\in(0.3-0.4)$&        $z \in (0.4-0.6)$&      
$z\in(0.6-0.8)$&  $\frac{dN^{K^{-}}(x)}{dN^{DIS}(x)}$ \\   \hline
0.033&  1.1& $0.195 \pm 0.015$ & $0.117 \pm 0.016$ & $ 0.060 \pm 0.005$ & $ 0.020 \pm 0.002$ & $  0.0472 \pm 0.0025$	  \\
0.048&  1.4& $0.181 \pm 0.006$ & $0.104 \pm 0.011$ & $ 0.057 \pm 0.004$ & $ 0.014 \pm 0.002$ & $  0.0429 \pm 0.0015$	  \\
0.065&  1.6& $0.170 \pm 0.013$ & $0.101 \pm 0.005$ & $ 0.050 \pm 0.004$ & $ 0.013 \pm 0.001$ & $  0.0398 \pm 0.0016$	  \\
0.087&  1.7& $0.134 \pm 0.019$ & $0.085 \pm 0.009$ & $ 0.041 \pm 0.003$ & $ 0.013 \pm 0.002$ & $  0.0327 \pm 0.0022$	  \\
0.120&  2.1& $0.134 \pm 0.021$ & $0.080 \pm 0.005$ & $ 0.037 \pm 0.003$ & $ 0.011 \pm 0.001$ & $  0.0309 \pm 0.0023$	  \\
0.170&  3.1& $0.135 \pm 0.027$ & $0.077 \pm 0.007$ & $ 0.033 \pm 0.003$ & $ 0.009 \pm 0.001$ & $  0.0296 \pm 0.0029$	  \\
0.240&  4.9& $0.134 \pm 0.015$ & $0.074 \pm 0.005$ & $ 0.029 \pm 0.003$ & $ 0.007 \pm 0.001$ & $  0.0279 \pm 0.0017$	  \\
0.340&  7.4& $0.142 \pm 0.021$ & $0.080 \pm 0.011$ & $ 0.031 \pm 0.005$ & $ 0.007 \pm 0.001$ & $  0.0300 \pm 0.0026$	  \\
0.450&  10.1&$0.176 \pm 0.054$ & $0.115 \pm 0.037$ & $ 0.031 \pm 0.007$ & $ 0.010 \pm 0.003$ & $  0.0372 \pm 0.0068$	  \\
\end{tabular}
\end{center}
\end{table}

\begin{table}[h!]
\begin{center}
\caption{Mean values of $x$ and $Q^2$ in (GeV/c)$^2$ for the used data as well as values of the $x \cdot$PDF 
for quark flavours from  MSTW08L. In the last column $4(u_v+d_v)/(5(u+ \bar{u} + d +\bar{d} ) +4s)$ is given. }
\label{tab:3}
\begin{tabular}{c|c|c|c|c|c|c|c}
  $\langle x \rangle$ &  $\langle Q^2 \rangle$  & $xu$ & $x\bar{u}$ & $xd$ & $x\bar{d}$ & $xs$ &  $\frac{4( u_v+ d_v)}{5(u+ 
\bar{u} + d +\bar{d} ) +4s}$ \\ \hline
0.033&  1.1&    0.408&  0.165&  0.346&  0.178&  0.063&  0.286 \\
0.048&  1.4&    0.450&  0.153&  0.359&  0.178&  0.062&  0.322 \\
0.065&  1.6&    0.488&  0.138&  0.367&  0.176&  0.057&  0.357 \\
0.087&  1.7&    0.527&  0.119&  0.372&  0.169&  0.050&  0.399 \\
0.120&  2.1&    0.579&  0.093&  0.375&  0.150&  0.040&  0.462 \\
0.170&  3.1&    0.624&  0.064&  0.358&  0.108&  0.027&  0.551 \\
0.240&  4.9&    0.618&  0.039&  0.302&  0.053&  0.015&  0.646 \\
0.340&  7.4&    0.513&  0.019&  0.202&  0.014&  0.006&  0.727 \\
0.450&  10.1&   0.344&  0.006&  0.104&  0.002&  0.002&  0.767 \\

\end{tabular}
\end{center}
\end{table}

\begin{table}[h!]
\begin{center}
\caption{As Table \ref{tab:3} but for CTEQ6L.}
\label{tab:4}
\begin{tabular}{c|c|c|c|c|c|c|c}
  $\langle x \rangle $&  $\langle Q^2 \rangle$  & $xu$ & $x\bar{u}$ & $xd$ & $x\bar{d}$ & $xs$ &  $\frac{4( u_v+ d_v)}{5(u+ 
\bar{u} + d +\bar{d} ) +4s}$ \\ 
\hline
0.033&	1.1&	0.423&	0.147&	0.324&	0.169&	0.063&	0.310 \\
0.048&	1.4&	0.465&	0.132&	0.344&	0.160&	0.058&	0.360 \\
0.065&	1.6&	0.505&	0.117&	0.361&	0.152&	0.054&	0.405 \\
0.087&	1.7&	0.549&	0.100&	0.376&	0.142&	0.048&	0.453 \\
0.120&	2.1&	0.601&	0.078&	0.386&	0.124&	0.042&	0.513 \\
0.170&	3.1&	0.637&	0.052&	0.367&	0.091&	0.030&	0.588 \\
0.240&	4.9&	0.616&	0.031&	0.305&	0.048&	0.017&	0.665 \\
0.340&	7.4&	0.498&	0.014&	0.202&	0.014&	0.006&	0.732 \\
0.450&	10.1&	0.326&	0.005&	0.106&	0.002&	0.002&	0.773 \\
\end{tabular}
\end{center}
\end{table}

\end{document}